\documentclass[12pt]{article}
\usepackage{amsmath,amssymb,exscale}
\usepackage{psfrag,graphicx}
\input epsf

\def\gappeq{\mathrel{\rlap {\raise.5ex\hbox{$>$}}
{\lower.5ex\hbox{$\sim$}}}}
\def\lappeq{\mathrel{\rlap{\raise.5ex\hbox{$<$}}
{\lower.5ex\hbox{$\sim$}}}}
\newcommand{\Tr}{\mathop{\rm Tr}}
\def\I{\rm 1\kern-.24em l}  

\begin{document}
\topmargin 0cm
\oddsidemargin 0cm

\pagestyle{empty}
\begin{flushright}
UAB-FT-588\\
October 2005
\end{flushright}
\vspace*{5mm}

\begin{center}
\vspace{2.5cm}
{\Large\bf
 The scalar and pseudoscalar sector
 in a \\ \vskip4mm
 five-dimensional approach
to chiral symmetry breaking
  }\\
\vspace{2.5cm}
{\large Leandro  Da Rold and
Alex Pomarol}\\
\vspace{.6cm}
{\it { IFAE, Universitat Aut{\`o}noma de Barcelona,
08193 Bellaterra, Barcelona}}\\
\vspace{.4cm}
\end{center}

\vspace{1cm}
\begin{abstract}
We study the scalar and pseudoscalar sector in  
a five-dimenional model describing  chiral symmetry breaking.
We calculate  the scalar and pseudoscalar
 two-point correlator,
 the mass  spectrum and interactions.
We also  obtain the scalar and pseudoscalar 
contributions   to the  coefficients of the chiral lagrangian
and  determine the scalar form factor of the pseudo-Goldstone bosons.
Most quantities show a good agreement with QCD.
\end{abstract}

\vfill
\eject
\pagestyle{empty}
\setcounter{page}{1}
\setcounter{footnote}{0}
\pagestyle{plain}
%

\section{Introduction}

The  AdS/CFT correspondence    \cite{adscft} has provided
a new approach to tackle  strongly coupled theories.
This has recently boosted the search for
 string realizations of   QCD-like theories \cite{flavoredstring,Sakai:2004cn}.
A crucial ingredient in these realizations is a (compact) warped extra dimension
that  plays the role of the energy scale in the QCD-like theory. 
It is therefore interesting  to look for  properties  of QCD in the strong regime
that can be  derived  from weakly coupled theories in  five-dimensions. 
Examples of this type of properties have already been  found in
high-energy  hadron  scattering \cite{Polchinski:2002jw}, string breaking \cite{Karch:2002xe}, 
hadron   form factors  and  hadron spectroscopy \cite{bbr}-\cite{evans}.

In Refs.~\cite{Erlich:2005qh,us}
a five-dimensional model was proposed to study the breaking of the chiral symmetry in QCD.
The model was described in terms  of infinite  weakly coupled resonances, similar to   QCD in the large-$N_c$ limit.
The vector sector was  studied  
and several relations among couplings and masses were  derived based only on the
 (warped) five-dimensionality of the space.
The predictions of the model showed  a  good agreement with   QCD.

Here we will extend this analysis to the scalar and pseudoscalar sector.
We will calculate the scalar and pseudoscalar two-point correlator and show 
that they have  a   similar behavior   to  that in QCD.
We will  obtain  the   masses and couplings of the resonances, pointing out
 the   implications  of working with a warped extra dimension.
We will also calculate the
scalar form factor of the pseudo-Goldstone boson (PGB)    
and the  pseudo(scalar) 
contributions to the   $L_i$ coefficients of the low-energy  chiral lagrangian.
A prediction for the  quark masses will also be given.
We will compare our results with 
 the QCD experimental data whenever this is available.

\section{A  five-dimensional model for QCD}
The 5D model   proposed to study the properties  of QCD with 3 flavors 
consists in a theory where  the  chiral
symmetry U(3)$_L\otimes$U(3)$_R$ is gauged in the 5D bulk 
\footnote{The U(1)$_A$ is broken by the anomaly that,
although it will not be studied here,  
can also be incorporated in extra-dimensional models  
along the lines of  Ref.~\cite{Sakai:2004cn}.}.
Parity is defined as
the interchange $L \leftrightarrow  R$. 
The bulk fields are the gauge bosons $L_M,R_M$ and a complex scalar field $\Phi$ 
transforming as a  (${\bf 3_L}$,${\bf\bar  3_R}$). 
This scalar plays the role of the operator $q\bar q$ in QCD  whose vacuum expectation
value (VEV) is responsible for the
breaking of the chiral symmetry.
The 5D metric in conformal coordinates is defined as
\begin{equation}
ds^2=a^2(z)\big(\eta_{\mu\nu}dx^\mu dx^\nu-dz^2\big)\, ,
\end{equation}
where $a$ is the warp factor. We will work within  AdS$_5$ 
\begin{equation}
a(z)=\frac{L}{z}\, , \label{ads}
\end{equation}
where  $L$ is the AdS curvature radius.  
The AdS$_5$ metric  will guarantee  conformal invariance of the model at high energies.
The fifth  dimension is assumed to be compact, $L_0\leq z\leq L_1$ \cite{Randall:1999ee}.
The boundary at $z=L_1$ generates  a  mass gap in the model 
(breaks the conformal symmetry at energies $\sim 1/L_1$),
while the boundary at $z=L_0$ is only needed to regulate the theory.
When performing  calculations one must take the limit $L_0\rightarrow 0$ and 
eliminate the  divergences  that one encounters by 
properly adding UV-boundary counterterms \cite{adscft}.
 The action is given by
\begin{equation}
S_5=\int d^4x\int dz\, {\cal L}_5\, ,
\end{equation}
where
\begin{equation}
{\cal L}_5 = \sqrt{g}\, {M_5} \Tr\left[-\frac{1}{4} {L_{MN}L^{MN}}
-\frac{1}{4}{R_{MN}R^{MN}} +\frac{1}{2}{|D_M\Phi|^2}
-\frac{1}{2}M^2_\Phi|\Phi|^2\right]\, .
\label{la5}
\end{equation}
The covariant derivative is defined as
\begin{equation}
D_M\Phi=\partial_M \Phi+iL_M\Phi-i\Phi R_M\, ,
\end{equation}
where $M=(\mu,5)$ and
$\Phi={\I}/\sqrt{3}\, \Phi_s+\Phi_aT_a$, 
 with  Tr$[T_aT_b]=\delta_{ab}$ (and similarly for $L_M$ and $R_M$).
 For  the value of the scalar mass we take
$M^2_\Phi=-3/L^2$ that, by the AdS/CFT dictionary, corresponds   
to associate the scalar $\Phi$ with a CFT operator
 of dimension 3 such as $\bar qq$.
  Solving
 the equation of
motion for $\Phi$ we obtain
\begin{equation}
\langle\Phi\rangle\equiv v(z)=c_1\, z+c_2\, z^3\, ,
\label{vev}
\end{equation}
where $c_1$ and $c_2$ 
can be written in terms of the  value of $v$
at the boundaries
\begin{equation}
c_1=\frac{\widetilde M_q L_1^3-\xi\,L_0^2}{LL_1(L_1^2-L_0^2)}\ ,\qquad
c_2=\frac{\xi-\widetilde M_qL_1}{LL_1(L_1^2-L_0^2)}\, ,
\end{equation}
where we have
defined
\begin{equation}
\widetilde M_q\equiv
\frac{L}{L_0}v\big|_{L_0}
\ ,\qquad \xi\equiv L\, v\big|_{L_1}\, .
\label{mqdef}
\end{equation}
It can be shown that a  nonzero $\widetilde M_q$ corresponds to an explicit breaking of the chiral symmetry in the UV,
while a nonzero $c_2$ corresponds to  a spontaneous breaking of the chiral symmetry  in the IR.
Therefore the value of $c_2$ is determined  dynamically by minimizing the action.
In order to get  a nonzero value for $c_2$ in the chiral limit 
($\widetilde M_q=0$)  
we add a potential for $\Phi$ on the IR-boundary:
\begin{equation}
{\cal L}_{\rm IR} = -a^4V(\Phi)\big|_{L_1} ,\qquad V(\Phi)=-\frac{1}{2}m^2_b\Tr|\Phi|^2+
\lambda\Tr|\Phi|^4\, .
\label{pot}
\end{equation}
An origin  for  this  type of potentials  can be found in string constructions 
\cite{flavoredstring,Sakai:2004cn}.
To determine the value of $c_2$, or equivalently the value of $\xi$,
we must minimize the effective 4D action obtained after  substituting  Eq.~(\ref{vev})  into the 5D action.
For $L_0\rightarrow 0$, this is given by
\begin{equation}
S_{\rm eff}\simeq -\int d^4x\Tr\left\{M_5L\left[\frac{-\widetilde M^2_q}{2L_0^2}+\frac{\widetilde M^2_q}{L_1^2}
-{2}\frac{\xi \widetilde M_q}{L_1^3}
+\frac{3}{2}\frac{\xi^2}{L_1^4}\right]+V(\xi)\frac{L^4}{L_1^4}\right\}\, ,
\label{energy}
\end{equation}
that is minimized  for
\begin{equation}
\label{xi}
\xi^2=\frac{\I}{4\lambda}\left(m_b^2L^2-3M_5L\right)+{\cal O}(\widetilde M_q)\, .
\end{equation}

This 5D model depends    on 5 parameters: 
\footnote{We  trade  $m^2_b$ for $\xi$ by means of Eq.~(\ref{xi}).
In the following we will take $\xi\rightarrow \xi{\I}+{\cal O}(\widetilde M_q)$ and treat $\xi$ 
as a parameter.}
$\widetilde M_q$, $M_5$, $L_1$,  $\xi$ and $\lambda$.
The value of $\widetilde M_q$ is related to the quark masses as we will see below. 
The values of $M_5$, $L_1$ and   $\xi$ 
were determined in Ref.~\cite{us}  from the gauge sector of the theory.
By using the QCD values for 
$N_c$, $M_\rho$ and $M_{a_1}$,  it was found \cite{us}:
\begin{equation}
M_5L=\frac{N_c}{12\pi^2}\equiv \tilde N_c ,\qquad
\frac{1}{L_1}\simeq 320\  {\rm MeV}  ,\qquad
\xi\simeq 4\, .
\label{fit}
\end{equation}
Our predictions will be given using the above values
 (although in certain cases we will  study the dependence of the predictions  on $\xi$).
This leaves
the scalar sector of the theory depending  only on one parameter, $\lambda$.  
An estimate of its value    can be obtained using naive dimensional analysis (NDA)
that   gives $\lambda\sim 1/(16\pi^2)\sim 10^{-2}-10^{-3}$.

\section{The scalar and pseudoscalar sector}

We define 
\begin{equation}
\Phi=\left(v+S\right) e^{iP/v}\, ,
\end{equation}
where  $S$ corresponds to a real scalar and $P$ to a real pseudoscalar under parity. 
Since we  will be  considering the chiral limit $\widetilde M_q\rightarrow0$,
we have $v\propto \I$  and  the  symmetry breaking pattern  
U(3)$_L\otimes$ U(3)$_R\rightarrow$ U(3)$_V$. 
Under SU(3)$_V$ we have that both $S$ and $P$   transform  as  ${\bf 1+8}$.
We  will work in the unitary gauge. 
This corresponds to add the gauge  fixing terms
\begin{equation}
\begin{aligned}
{\cal L}_{GF}^V&=-\frac{M_5
  a}{2\xi_V}\Tr\left[\partial_{\mu}V_{\mu}-\frac{\xi_V}{a}\partial_5
(a V_5)\right]^2\, ,
\\
{\cal L}_{GF}^A&=-\frac{M_5
  a}{2\xi_A}\Tr\left[\partial_{\mu}A_{\mu}-\frac{\xi_A}{a}\partial_5 (a
  A_5)-\xi_A \sqrt{2}a^2 v P\right]^2\, ,
\end{aligned}
\end{equation}
where
$V_M,A_M=\frac{1}{\sqrt{2}}\big(L_M\pm R_M)$,
and take the limit  $\xi_{V,A}\rightarrow\infty$, {\it i.e.}
\begin{equation}
\partial_5 (a V_5)=0\ ,\qquad  
P=-\frac{1}{\sqrt{2} a^3 v}\partial_5 (a A_5)\, .
\label{restric}
\end{equation}
The  above equation will allow us to write    $P$  
as a function of $A_5$ in the 5D lagrangian.
After integration by parts, the  5D quadratic terms for the scalar $S$ and the
pseudoscalar $A_5$  are given by
\begin{equation}\label{L2}
\begin{aligned}
&{\cal L}_S=-\frac{a^3 M_5}{2} \Tr\Big\{
S[\partial^2-a^{-3}\partial_5a^3\partial_5
+a^2 M_{\Phi}^2]S\Big\},
\\
&{\cal L}_{A_5}=-\frac{aM_5}{2} \Tr\Big\{A_5\big[\partial^2
{\cal D}+{\cal D}(2v^2a^2{\cal D})\big]A_5\Big\},
\end{aligned}
\end{equation}
\noindent where ${\cal D}$ is a differential operator defined by
\begin{equation}
{\cal D}=1-\partial_5\left(\frac{1}{2v^2a^3}\partial_5 a\right)\, .
\end{equation}
The  scalar  and pseudoscalar field  has also  4D boundary terms  that, 
after using the 5D equation of motion
 for $A_5$ 
({\it i.e.},  ${\cal D}A_5=-\partial^2A_5/(2v^2a^2)$)
 and Eq.~(\ref{xi}),   can be written as 
\footnote{One obtains  the same result if,
instead of the equation of motion,
one uses    the mass eigenfunction equation,  
 ${\cal D}A_5=m^2 A_5/(2v^2a^2)$, as we will do later to perform a Kaluza-Klein
  decomposition
 of the sector.}
\begin{eqnarray}
\label{Lbound}
{\cal L}_{bound}&=&
-\frac{M_5a}{2}\Tr\big[a^2 S\partial_5 S
+A_5\frac{\partial^2}{2v^2a^3}\partial_5(aA_5)+2A_\mu\partial_\mu A_5\big]\Big{|}_{L_0}^{L_1}
-a^4V(S)\big|_{L_1}\nonumber\\
&+&M_5a^3\Tr[ S]\partial_5 v\big|_{L_0}\, ,
\end{eqnarray}  
where
\begin{equation}\label{VIR}
V(S)\big|_{L_1}=m^2_{S}\Tr[S^2]\big|_{L_1}+{\cal O}(S^3) ,\qquad
m^2_S=\frac{4\lambda\xi^2}{L^2}-\frac{3M_5}{2L}+
{\cal O}(\widetilde M_q)\, .
\end{equation}
To cancel the quadratic terms on the IR-boundary of Eq.~(\ref{Lbound}) we impose the conditions
\begin{equation}\label{irbc}
\big[M_5\partial_5+2a m^2_S\big]S\big{|}_{L_1}=0 ,\qquad A_5\big{|}_{L_1}=0\, .
\end{equation}
The boundary conditions on the UV-boundary  will be specified later.

The interactions between scalars and pseudoscalars that we will be considering are
\begin{eqnarray}\label{LSPP}
{\cal L}_{SA_5A_5}&=&\frac{a^3M_5}{2}\Tr\left[\frac{S}{v^3a^6}\Big{(}\partial_{\mu}
\partial_5(aA_5)\Big{)}^2-4vS({\cal D}A_5)^2\right]\, ,
\\
\label{LP4}
{\cal L}_{A_5^4}&=&\frac{M_5}{96a^9
  v^6}\Tr\Big[\big(\partial_5(aA_5)\overleftrightarrow{\partial_{\mu}}\partial_5(aA_5)\big)^2\Big]\, .
\end{eqnarray}
The $SVV$ interaction is absent. 
This is a  consequence of the  U(3)$_V$  invariance 
and the fact that only  dimension-four  operators are considered in Eq.~(\ref{la5}).
 This interaction, however,  could be
induced by higher-dimensional operators or loop effects.

With the above lagrangian for the scalar and pseudoscalar sector   
we can  calculate any relevant physical quantity. 
We will be considering two approximations.
First, we will be working at the tree-level. 
According to Eq.~(\ref{fit})
this corresponds to
work in the large-$N_c$ limit.
Since loop effects are expected to be of order $1/N_c$,
 our predictions for QCD quantities will have a   $30\%$ uncertainty.
Second,  
we will take the chiral limit $\widetilde  M_q\rightarrow 0$.
For the pseudoscalar sector this limit will be taken  in the following way.
We will first perform the calculations with  $c_1\rightarrow 0$ and fixed $L_0$ 
 (this is equivalent to $\widetilde M_q\rightarrow \xi L_0^2/L_1^3$ and
$c_2\rightarrow \xi/(LL_1^3$)).  Next we will take  the limit $L_0\rightarrow 0$. 
This procedure 
simplifies the calculations and 
avoids singularities at $z=L_0$.

\subsection{The scalar and pseudoscalar  correlator}

In this section we will calculate the  scalar and pseudoscalar two-point correlator.
In  QCD these are defined as
\begin{equation}
\Pi_{S,P}(p^2)=-\int d^4x e^{ipx}\langle J_{S,P}(x)J_{S,P}(0)  \rangle\, ,
\end{equation}
where $J_S=\bar q q$ and $J_P=i\bar q\gamma_5 q$. 
The correlators $\Pi_{S,P}$ can be obtained from the  generating functional ${\cal S}$
according to
\begin{equation}  
\Pi_S=\frac{\delta^2 {\cal S}}{\delta s^2}\ ,\qquad
\Pi_P=\frac{\delta^2 {\cal S}}{\delta p_s^2}\, ,
\end{equation}
where $s$ and  
$p_s$ are the scalar and pseudoscalar  external sources coupled to QCD:
\begin{equation}
{\cal L}=-\Tr[ \bar q_L\, \phi\,  q_R]+h.c. ,\quad \phi=M_q+s+ip_s\, .
\end{equation}
The AdS/CFT correspondence tells us \cite{adscft}  that
${\cal S}$ is obtained in the 5D theory 
by integrating out the bulk fields restricted to a given UV-boundary value.
These boundary values play the role of the external sources coupled to  QCD. 
In particular, for the 5D    scalar field  we have 
\begin{equation}
\label{uvbc}
\Phi\big|_{L_0}=\alpha\frac{L_0}{L} \phi\, ,
\end{equation}
where the constant $\alpha$  will  be determined by matching with the QCD correlators in the UV
as we will see later.
Up to the quadratic order  in the fields, Eq.~(\ref{uvbc}) leads to 
\begin{equation}
\label{uvbcb}
S\big|_{L_0}=\alpha \frac{L_0}{L}\left(s+\alpha\frac{p^2_s}{2\widetilde M_q}\right) 
,\qquad P\big|_{L_0}=-\frac{\partial_5(aA_5)}{\sqrt{2}a^3v}\Big|_{L_0}=\alpha \frac{L_0}{L}p_s\, .
\end{equation}

Let us calculate ${\cal S}=\int d^4x\,  {\cal L}_{\rm eff}$   at the quadratic  level
for $S$ and $A_5$. By solving
 the equations of motion from Eq.~(\ref{L2}) with the boundary
conditions of Eqs.~(\ref{irbc}) and
(\ref{uvbcb}), and substituting the solution back into the action, 
we get (in momentum space)
\footnote{There is  also a  mixing term between  $p_s$ and the longitudinal part of  $A_\mu|_{L_0}$ 
that we are not writing.}
\begin{equation}\label{Leff}
{\cal L}_{\rm eff}
=\frac{1}{2}\Pi_S(p^2)\Tr[s^2]+\frac{1}{2}\Pi_P(p^2)\Tr[p_s^2] + \Gamma_S\Tr[s]\, .
\end{equation}
For a AdS$_5$ space $\Pi_S$ can be given analytically at the tree-level. We obtain
\begin{equation}
\label{PiS}
\Pi_S(p^2)=\alpha^2 M_5L\left[\frac{1}{L_0^2}+ \frac{ip}{L_0}\frac{J_0(ip L_0)
+b(p) Y_0(ip L_0)} {J_1(ip L_0)+b(p)Y_1(ip L_0)}\right] \, ,
\end{equation}
where $J_n,Y_n$ are Bessel functions, $p$ is the
Euclidean momentum and  $b(p)$ is determined by the IR-boundary condition of Eq.~(\ref{irbc}):
\begin{equation}
\label{bS}
b(p)=-\frac{ipL_1J_2(ipL_1)-\frac{8\lambda\xi^2}{M_5L}J_1(ipL_1)}{ipL_1Y_2(ipL_1)-\frac{8\lambda\xi^2}{M_5L}Y_1(ipL_1)}\, .
\end{equation}
Taking the limit $L_0\rightarrow 0$ we find
\begin{equation}
\label{pis2}
\Pi_S(p^2)\simeq
\alpha^2M_5L\left[\frac{1}{L_0^2}+\frac{1}{2}p^2
  \ln(p^2L_0^2)+\frac{\pi p^2}{2 b(p)}\right]\, .
\end{equation}
The divergent terms  for $L_0\rightarrow 0$ can be absorbed in a bare mass and a bare kinetic term for $s$.
After this renormalization the correlator is finite.
For large  momentum $pL_1\gg 1$, we find
\begin{equation}\label{largepS}
\Pi_S(p^2)\simeq\frac{\alpha^2M_5L}{2}\, p^2 \ln p^2\, ,
\end{equation}
as  expected from the conformal symmetry.  
Matching with QCD   in which  at large momentum  we have
\begin{equation}
\Pi^{\rm QCD}_S(p^2)\simeq\frac{N_c}{8\pi^2}\, p^2 \ln p^2\, ,
\end{equation}
we obtain, using Eq.~(\ref{fit}),
\begin{equation}
\label{alpha}
\alpha=\sqrt{3}\, .
\end{equation}
The next to leading terms in the large momentum expansion in Eq.~(\ref{largepS}) are suppressed exponentially,
contrary to  QCD where one finds  that the scalar correlator has power corrections. 
This is because we assumed, for simplicity, 
 that the scalar had a potential only  on the IR-boundary.
In more realistic models  such as those arising from  string theories 
the scalar potential is present    in the 5D bulk (although  peaked towards the IR).
In these cases     the scalar correlator  has   power corrections.
Also, if the 5D metric deviate in the IR from AdS or if higher-dimensional operators are
included in Eq.(\ref{la5}), then  power corrections can be present
in $\Pi_S$.

\begin{figure}[t]
  \centering
  \psfrag{m}{\hskip-5mm Mass [MeV]}
  \psfrag{mS1}{$S_1$}
  \psfrag{mS2}{$S_2$}
  \psfrag{a0980}{$a_0(980)$}
  \psfrag{a01450}{$a_0(1450)$}
  \psfrag{lambda}{$\lambda\, [10^{-3}]$}
  \includegraphics[width=8.5cm]{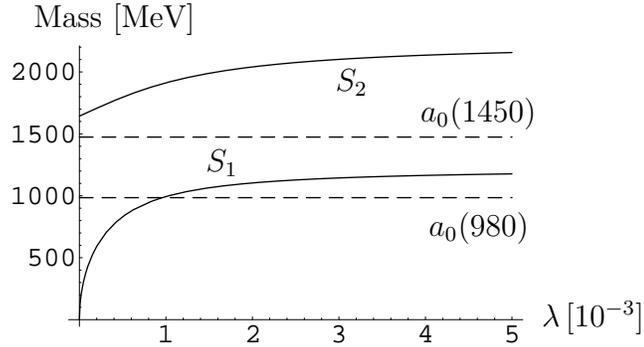}
  \caption{\textit{Mass of the first  and second scalar resonances
as a function of $\lambda$ for $\xi=4$. The dashed lines show the experimental values
for the  masses of the scalar resonances   $a_0(980)$ and $a_0(1450)$.}}
  \label{figmS}
\end{figure}

For small momentum $\Pi_S(p^2)$ can be approximated by
\begin{equation}\label{smallpS}
\Pi_S(p^2)\simeq 3\tilde{N}_c\,\left[-\frac{2}{L_1^2}+\frac{\tilde N_c}{2\lambda\xi^2L_1^2}\right]+
{\cal O}(p^2)\, .
\end{equation}
The scalar correlator Eq.~(\ref{PiS}) can also  be written as a sum over infinitely narrow resonances, similarly
as in large-$N_c$ QCD:
\begin{equation}\label{sigma}
\Pi_S(p^2)=\sum_n \frac{F_{S_n}^2M^2_{S_n}}{p^2+M_{S_n}^2} \, .
\end{equation}
Therefore the  masses of the scalar resonances  can be  determined by finding  the poles of Eq.~(\ref{pis2}),
{\it i.e.}, by the equation $b(p)=0$.
In Fig.~\ref{figmS} we plot the value of the mass of the
first and second scalar resonance as a function of $\lambda$ for $\xi=4$.
The first resonance mass ranges from $M_{S_1}=0$  MeV ($\lambda\rightarrow 0$) to 
$M_{S_1}=1226$ MeV   ($\lambda\rightarrow\infty$). 
We  compare this  value with the masses of the  $a_0$ states 
(since these are the QCD scalars whose masses are not very sensitive to  $M_q$).
We see that for a value of $\lambda$ close to its NDA estimate, $\lambda\sim 10^{-2}-10^{-3}$,
the  mass of the first scalar resonance is closer to that of  $a_0(980)$  than to that of $a_0(1450)$.
Nevertheless we must recall that we are working in the large-$N_c$ limit and
then corrections can be as large as $30\%$. 
Consequently  we cannot discard to associate $S^{(1)}$ with $a_0(1450)$. 
The scalar decay constants $F_{S_n}$ are determined by the residues of $\Pi_S$. We obtain
\begin{equation}
F^2_{S_n}=
\frac{3\tilde N_c 
\pi M_{S_n}^2\left(
\frac{8\lambda\xi^2}{M_5L}Y_1(M_{S_n}L_1)
-M_{S_n}L_1Y_2(M_{S_n}L_1)\right)}
{M_{S_n}L_1\left(1-\frac{8\lambda\xi^2}{M_5L}\right)J_0(M_{S_n}L_1)+
  \left(\frac{8\lambda\xi^2}{M_5L}+M_{S_n}^2L_1^2-2\right)J_1(M_{S_n}L_1)}\, .
\end{equation}
For $\lambda\simeq 10^{-3}$ we obtain $M_{S_1}\simeq 1$ GeV and 
 $F_{S_1}\simeq 260$ MeV,
while for  the second resonance  we get
 $M_{S_2}\simeq 1900$ MeV and  $F_{S_2}\simeq 370$ MeV.
Using this result we can calculate  the value of the coupling $c_m$ defined in 
Ref.~\cite{Ecker:1988te}. We obtain
$c_m=F_{S_1}M_{S_1}/(4B_0)\simeq 41$ MeV  (taking  the value of $B_0$ 
from Eq.~(\ref{b0}))
 very close to the value used in
 Ref.~\cite{Ecker:1988te}:   $c_m\simeq42$ MeV.

To calculate the pseudoscalar correlator $\Pi_P$   we must  rely on  numerical analysis.
Only for small
and large momentum  we are able to give analytical results.
For large momentum $pL_1\gg 1$ 
we have
\begin{equation}\label{largepP}
\Pi_P(p^2)=\frac{3\tilde N_c}{L_0^2}+p^2\left[
\frac{3\tilde N_c}{2} \ln(p^2L_0^2)-\frac{c^P_6}{p^6}
+{\cal O}(\frac{1}{p^{12}})\right]\ ,
\qquad {\rm where}\ \ \
c^P_6=-\frac{64}{5}\frac{3\tilde N_c\, \xi^2}{L_1^6}
\, .
\end{equation}
Again the divergences can be cancelled  by adding a proper 
  mass and a kinetic term  for the pseudoscalar $p_s$ on the UV-boundary.
From Eqs.~(\ref{largepS}) and (\ref{largepP})
 we can  obtain 
the correlator $\Pi_{SP}=\Pi_S-\Pi_P$   at large momentum.
It  drops as  $\Pi_{SP}\sim c^P_6/p^4$.
Comparing with $\Pi_{LR}=\Pi_V-\Pi_A\sim c_6/p^4$, we find $c^P_6=12\, c_6$ \cite{us}
 in strong disagreement with QCD in which one has $c^P_6=3\, c_6$.
This can be improved if, as we said, we consider more realistic theories where the scalar
potential is present  in the 5D bulk and therefore  $\Pi_S$ has power corrections.

At low momentum \footnote{In order to obtain the correct result it is important to take    the limit $L_0\rightarrow 0$ before   
taking $p^2\rightarrow 0$ \cite{Contino:2004vy}.} 
and for $\xi\gg 1$ we find
\begin{equation}\label{smallpP}
\Pi_{P}(p^2)\simeq 
  \frac{2\widetilde B^2_0 F^2_\pi}{p^2}-\tilde N_c\widetilde B_0^2
 + {\cal O}(p^2)\, ,
 \end{equation}
where
\begin{equation}
\label{fpi}
 \widetilde B_0=\frac{2\sqrt{3}\tilde{N}_c \xi}{F_{\pi}^2 L_1^3}\ ,\qquad
F_\pi^2=\Pi_A(0)\stackrel{\xi\gg 1} {\simeq}
\frac{2^{5/3}\pi\tilde N_c}{3^{1/6}\Gamma(\frac{1}{3})^2}
\frac{\xi^{2/3}}{L_1^2}
\, .
\end{equation}
$\Pi_A(p^2)$ is the axial-vector correlator calculated in Ref.~\cite{us}.
The first term of Eq.~(\ref{smallpP})  shows a pole at $p^2=0$ as expected due to the presence of the 
 massless PGB.

\begin{figure}[t]
  \centering
  \psfrag{mP1}{$M_{P_1}$[MeV]}
  \psfrag{x}{$\xi$}
 \includegraphics[width=8.5cm]{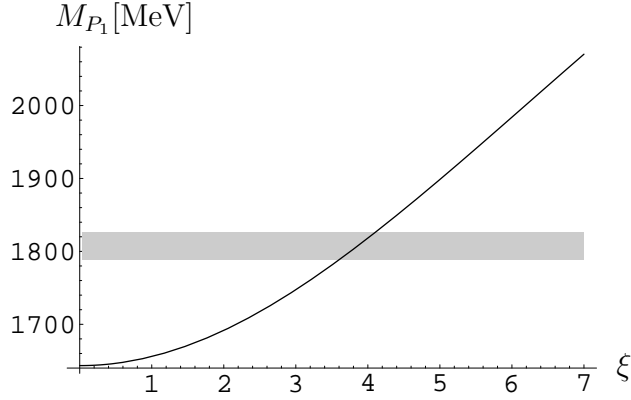}
  \caption{\textit{Mass of the first massive  pseudoscalar resonance
as a function of $\xi$. The shadow band shows  the experimental value
for $\pi(1800)$.}}
  \label{figmP}
\end{figure}

By looking at the poles of $\Pi_P$ we can find the 
pseudoscalar masses.
The lowest mode is the 
massless  PGB of the
spontaneous chiral symmetry breaking.
There  is a nonet of PGBs but we must recall that the inclusion of the  U(1)$_A$-anomaly 
will give mass to the singlet \cite{Sakai:2004cn}.
The mass of the first massive resonance is shown in Fig.~\ref{figmP}
as a function of~$\xi$.
We see that its value is far from the mass of the $\pi(1300)$ state. 
Nevertheless,  we find that, for $\xi\simeq 4$, $M_{P_1}$   is close to the mass of 
 $\pi(1800)$ suggesting that this could be the state to be associated with our first massive pseudoscalar  resonance.  For this resonance we find a decay constant $F_{P_1}\simeq 374$ MeV.

Finally, we calculate   the linear term in Eq.~(\ref{Leff}) 
to be associated  in QCD with the    $\bar qq$ condensate: $\Gamma_S=-\langle J_S\rangle$. 
We find
\begin{equation}
\label{cond}
\Gamma_S=
\sqrt{3} \tilde N_c\frac{\widetilde M_qL_1^2+2 \xi L_0^2/L_1-3\widetilde M_q L_0^2}{L_0^2 (L^2_1-L_0^2)}
\ \ \ \stackrel{\widetilde M_q\to 0} {\longrightarrow }\ \ \  \frac{2\sqrt{3} 
\tilde N_c\, \xi}{L_1 (L_1^2-L_0^2)}
\ \ \ \stackrel{L_0\to 0} {\longrightarrow }\ \ \  \frac{2\sqrt{3} 
\tilde N_c\, \xi}{L_1^3}\, .
\end{equation}

\subsection{Scalar meson interactions}
\begin{figure}[t]
  \centering
  \psfrag{S1}{\small{$S_1$}}
  \psfrag{S2}{\small{$S_2$}}
  \psfrag{P1}{\small{$P_1$}}
  \psfrag{P0}{\small{$\pi$}}
  \psfrag{z}{\Large{$\frac{z}{L_1}$}}
   \includegraphics[width=8.5cm]{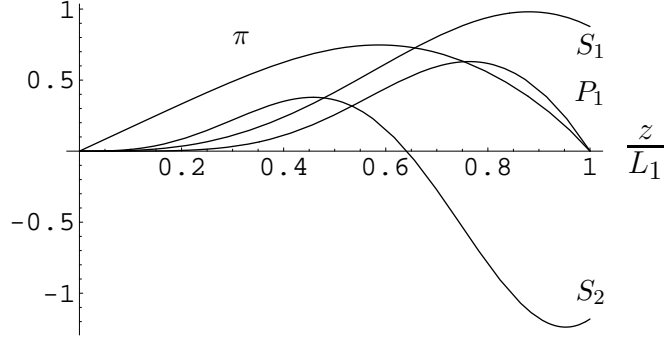}
  \caption{\textit{Wave-functions of  the $n=1,2$ scalar
  resonance,  the PGB and the first massive pseudoscalar for $\xi=4$
  and $\lambda=10^{-3}$.}}
  \label{figKKmodes}
\end{figure}

To study the interactions it is convenient to perform
a Kaluza-Klein (KK) decomposition of the 5D fields:
 \begin{equation}
 S(x,z)= \frac{1}{\sqrt{M_5L}}\sum_{n=1}^\infty f^S_{n}(z)S^{(n)}(x)\ ,\ \ \ \ 
  A_5(x,z)=\frac{1}{\sqrt{M_5L}}\sum_{n=0}^\infty f^P_{n}(z)P^{(n)}(x)\, .
  \end{equation}
 We impose the following boundary conditions on the UV-boundary:
\begin{equation}\label{KKuvbc}
S\big{|}_{L_0}=0 ,\qquad
P\big|_{L_0}\propto
\partial_5(aA_5)\big|_{L_0}=0\, ,
\end{equation}
that  cancel the boundary terms of Eq.~(\ref{Lbound}).
The  wave-functions of the KK-modes $S^{(n)}$
are given by
\begin{equation}
f_n^{S}(z)=\frac{z^2}{N_{S_n}L_1^2}\left[ J_1(M_{S_n}z)-\frac{J_1(M_{S_n}L_0)}{Y_1(M_{S_n}L_0)}
Y_1(M_{S_n}z)\right] \ \ \ \stackrel{L_0\to 0} {\longrightarrow }\ \ \
 \frac{z^2}{N_{S_n}L_1^2}
J_1(M_{S_n}z)
\, , \label{wfv}
\end{equation}
where $N_{S_n}$ is a constant fixed by canonically normalizing the
fields, $\int a^3 (f^{S}_n)^2 dz/L=1$.
In Fig.~\ref{figKKmodes}
we plot
the wave-functions of the first two KK-modes.

The equation that determines the wave-functions of the pseudoscalars  can be obtained from Eq.~(\ref{L2}).
This is given by
\begin{equation}\label{KKP}
{\cal D} f^{P}_n=\frac{M_{P_n}^2}{2v^2a^2}f^{P}_n\, .
\end{equation}
The lowest state, $P^{(0)}\equiv\pi$,  is the PGB that in the limit   $L_0\rightarrow 0$ is massless. 
Its wave-function is given by
\begin{equation}
f^{\pi}(z)\ \ \ \stackrel{L_0\to 0} {\longrightarrow }\ \ \
\frac{z^3}{L_1^3 N_0}\left[I_{2/3}\left(\frac{\sqrt{2}\xi}{3}
\frac{z^3}{L_1^3}
\right)-
\frac{I_{2/3}\left(\sqrt{2}\xi/3\right)}{K_{2/3}\left(\sqrt{2}\xi/3\right)}
K_{2/3}\left(
\frac{\sqrt{2}\xi}{3}
\frac{z^3}{L_1^3}
\right)\right] \, ,
\end{equation}
where   $N_0$ 
is  determined  by the condition $-\frac{1}{2a^2v^2L}f^{\pi}\partial_5 (af^{\pi})|_{L_0}=1$.
The wave-function of the massive modes must be obtained numerically
from Eq.~(\ref{KKP}) with the 
normalization condition
$\int dz\,(f^{P}_n M_{P_n})^2/(2v^2aL)=1$.
The wave-functions of  $\pi$  and $P^{(1)}$ are shown in Fig.~\ref{figKKmodes}.

\begin{figure}[t]
  \centering
    \psfrag{gn}{\hskip-5mm\small{$G_{n\pi\pi}\, [{\rm GeV}^{-1}]$}}
  \psfrag{g1}{\small{$G_{1\pi\pi}$}}
  \psfrag{g2}{\small{$G_{2\pi\pi}$}}
  \psfrag{g3}{\small{$G_{3\pi\pi}$}}
  \psfrag{g4}{\small{$G_{4\pi\pi}$}}
  \psfrag{lambda}{$\lambda\, [10^{-3}]$}
    \includegraphics[width=8.5cm]{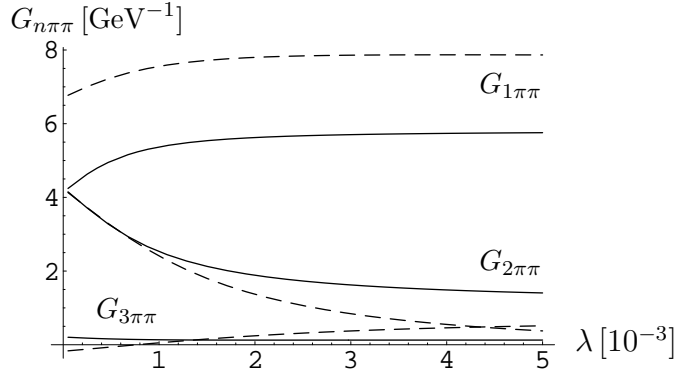}
  \caption{\textit{Coupling of the $n=1,2,3$ scalar resonance to two
  PGBs as a function of $\lambda$ for $\xi=4$ (solid line) and $\xi=3$ (dashed line).}}
  \label{figgn}
\end{figure}

The couplings  between the resonances are easily obtained by  integrating the 5D
interactions over $z$ with the corresponding wave-functions.
The coupling of a scalar to two PGBs comes from Eq.~(\ref{LSPP}). We obtain
\begin{equation}
{\cal L}_{S_n\pi\pi}=G_{n\pi\pi}\Tr[S^{(n)}(\partial_{\mu}\pi)^2]\, ,
\end{equation}
where $G_{n\pi\pi}$ is given by
\begin{equation}
G_{n\pi\pi}=\frac{1}{\sqrt{M_5L^3}}\int dz \,
f_n^S\frac{[\partial_5 (af^{\pi})]^2}{2a^3v^3}\, .
\end{equation}
In Fig.~\ref{figgn} we show the coupling of the first modes as a
function of $\lambda$ for $\xi=3,4$. 
We find that $G_{n\pi\pi}$ becomes smaller as $n$ increases.
This property is also present   in the  coupling between   a vector resonance and  two PGBs,
and it is due to the oscillatory behaviour of the KK wave-functions.
Associating  $S^{(1)}$ with  $a_0(980)$,  we find  that $M_{S_1}\simeq 980$ MeV
for $\lambda\simeq 10^{-3}$, and the prediction of the 5D  model for 
  the $a_0\pi\eta$  coupling is  $G_{1\pi\pi}\simeq 5.4$ GeV$^{-1}$
for $\xi=4$. 
In the notation of   Ref.~\cite{Ecker:1988te}
we find    $c_d=F^2_\pi G_{1\pi\pi}/2 \simeq 20$ MeV
to be compared to the value   $|c_d|\simeq 32$ MeV given there.
If  the width  of $a_0(980)$  is dominated by the decay to $\eta\pi$ we find
\begin{equation}
\Gamma(a_0\rightarrow \eta\pi)\simeq 27-56\ {\rm MeV}\ , \ \ \text{for}\ \xi=4-3\, .
\end{equation}
Unfortunately, the experimental value of the width of $a_0(980)$  has
 a large uncertainty   $\Gamma(a_0)=50-100$ MeV \cite{pdg}.

\section{(Pseudo)Scalar contributions to  PGB interactions}

By integrating the heavy scalar resonances we obtain the following  four-PGB interaction
\begin{equation}\label{L4pi8}
{\cal L}^{(8)}_{\pi^4}=\frac{1}{2}\left\{\Tr[(\partial_\mu\pi)^2 (\partial_\nu\pi)^2]
-\frac{1}{3}{\Tr}^2[(\partial_{\mu}\pi)^2]\right\}
  \sum_n\frac{G_{n\pi\pi}^2}{p^2+M_{S_n}^2}\, ,
\end{equation}
from  the scalar octet and 
\begin{equation}\label{L4pi1}
{\cal L}^{(1)}_{\pi^4}=
  \frac{1}{6}{\Tr}^2[(\partial_{\mu}\pi)^2]
  \sum_n\frac{G_{n\pi\pi}^2}{p^2+M_{S_n}^2}\, ,
\end{equation}
from  the scalar singlet.
The sum over the KK-modes in Eqs.~(\ref{L4pi8}) and (\ref{L4pi1})
 is dominated  by the first resonance.  
 At large momentum we find that the first resonance gives   $82\%$ of the total contribution
 and this percentage rises to $94\%$ at zero momentum (for $\lambda\simeq 10^{-3}$).
Therefore, as in the vector case \cite{us}, 
we find that the scalar mediation of the four-PGB interaction is  dominated
by the exchange of the first resonance.

Four-PGB interactions can also arise  from Eq.~(\ref{LP4}).  We find 
\begin{equation}\label{pion4}
{\cal L}_{\pi^4}=\frac{g_{\pi^4}}{4}\Tr[(\pi\overleftrightarrow{\partial_\mu}\pi)^2]\ ,\  \ \text{where}\ \ \ \ \
g_{\pi^4}=\frac{1}{24M_5L^2}\int  dz\,  \frac{[\partial_5(af^\pi)\big]^4}{a^9v^6}\, .
\end{equation}
At high energies  
the four-PGB amplitude arising from Eq.~(\ref{pion4}) grows as $\sim E^2$.
Nevertheless, this bad energy behavior of the four-PGB amplitude is 
cured by the contribution 
arising from Eqs.~(\ref{L4pi8}) and  (\ref{L4pi1}) that cancels the $E^2$ terms.
This occurs   thanks to the  sum rule
\begin{equation}
\label{sumrule1}
  \sum_n G_{n\pi\pi}^2=6\,  g_{\pi^4}\, .
\end{equation}
Eq.~(\ref{sumrule1}) is  a property of any 5D model in which the breaking of the chiral symmetry
is realized by the Higgs mechanism.

We can also calculate the coupling of the PGB to the source $s$ 
that defines the scalar form factor of the PGB.
Apart from a contact piece given by
\begin{equation}
\label{direct}
{\cal L}_{\pi^2s}=-\widetilde B_0\Tr[\pi^2s]\, ,
\end{equation}
this coupling is mediated by the
octet and singlet scalar resonances that gives respectively
\begin{eqnarray}
\label{sff8}
{\cal L}^{(8)}_{\pi^2s}&=&\left\{\Tr[(\partial_{\mu}\pi)^2s]-
  \frac{1}{3}{\Tr}[(\partial_{\mu}\pi)^2]\Tr[s]\right\}
\sum_n\frac{G_{n\pi\pi}F_{S_n}M_{S_n}}{p^2+M_{S_n}^2} \, ,\nonumber\\
 {\cal L}^{(1)}_{\pi^2s}&=&
  \frac{1}{3}{\Tr}[(\partial_{\mu}\pi)^2]\Tr[s]
  \sum_n\frac{G_{n\pi\pi}F_{S_n}M_{S_n}}{p^2+M_{S_n}^2}\, .
\end{eqnarray}
The scalar form factor of the PGB 
is then given by (normalized to unity  at zero momentum)
\begin{equation}
\label{fspi}
{\cal F}^S_\pi(p)= 1-\frac{p^2}{2\widetilde B_0}
\sum_n\frac{G_{n\pi\pi}F_{S_n}M_{S_n}}{p^2+M_{S_n}^2}\, .
\end{equation}
At low momentum the sum in Eq.~(\ref{fspi})   is dominated 
by the first resonance  that gives  $75\%$
of the total contribution (for $\lambda\simeq 10^{-3}$).
At large momentum we find that
the form factor goes as  $1/p^2$, as expected from  the conformal symmetry 
 \cite{Polchinski:2002jw}.
The cancellation of the constant term in ${\cal F}^S_\pi(p)$ occurs   due to the sum rule
\begin{equation}
\label{sumrule2}
\sum_n G_{n\pi\pi}F_{S_n}M_{S_n}=2\widetilde  B_0\, .
\end{equation}
This sum rule is fulfilled in any 5D model whose metric approaches to AdS$_5$ 
for $z\rightarrow 0$ (conformal theories in the UV).
In Eq.~(\ref{sumrule2})  we find that the first two resonances give a similar contribution,
while the contributions of the heavier resonances tend to cancel out. 
 Therefore we   see that ${\cal F}^S_\pi(p)$ is 
very well approximated by  the  exchange of only the first two resonances.

\subsection{The Chiral Lagrangian}

At energies below the massive resonances  our  5D model  is  described by the QCD chiral lagrangian.
In this section we calculate the (pseudo)scalar contributions 
to the   coefficients  of the chiral lagrangian
and compare these results with the QCD values.

Up to   ${\cal O} (p^2)$, the chiral lagrangian 
for the  octet of PGB, $\pi=\pi_aT_a$, 
 is given by \cite{Gasser:1984gg}
\begin{equation}
{\cal L}_2
=\frac{F_\pi^2}{4}\,
\Tr\big[ D_\mu U^\dagger D^\mu U \, + \, U^\dagger\chi  \,
+  \,\chi^\dagger U\big]\, ,
\label{l2}
\end{equation}
where
\begin{equation}
D_\mu U = \partial_\mu U - i R_\mu U + i U L_\mu ,\qquad   U= e^{i\sqrt{2}\, {\pi}/{F_\pi}}\, , 
 \end{equation}
and
\begin{equation}
\chi  = \, 2 B_0 \, ( M_q +s+ i p_s)\ ,\qquad  M_q={\rm Diag}(m_u,m_d,m_s)\, .
\label{chi}
\end{equation}
The prediction of our model for $F_\pi$  is given in Eq.~(\ref{fpi}). It gives
\begin{equation}
\label{fpiv}
F_\pi
\simeq  87 \left(\frac{\xi}{4}\right)^{\frac{1}{3}}\ {\rm MeV}
\, .
\end{equation}
For the prediction of   $B_0$ we can use Eq.~(\ref{cond}):
\begin{equation}
\label{ b01}
\langle \bar q q\rangle=-F^2_\pi B_0=-2\sqrt{3}\tilde N_c\frac{\xi}{L_1^3}\simeq 
-(226\ {\rm MeV})^3\left(\frac{\xi}{4}\right)\, ,
\end{equation}
that leads to
\begin{equation}
\label{b0}
B_0=\frac{2\sqrt{3}\tilde N_c\xi}{F^2_\pi L_1^3}\simeq 1520\left(\frac{\xi}{4}\right)^{\frac{1}{3}} \ {\rm MeV}\, .
\end{equation}
Notice that $B_0=\widetilde B_0$ as it should be, since   the  first term of Eq.~(\ref{smallpP}) 
can also be deduced by integrating out 
the PGB at tree-level in the chiral lagrangian. 
The relation $B_0=\widetilde B_0$ also leads to the 
right matching of Eq.~(\ref{direct})  with the chiral lagrangian.
The value of the quark masses $M_q$ is related to the VEV of $\Phi$ on the UV-boundary. 
Using   Eqs.~(\ref{mqdef}), (\ref{uvbc})  and (\ref{alpha}) we obtain
\footnote{In Refs.~\cite{Erlich:2005qh, us} the quark masses did not have the correct normalization since the 
value of  $\alpha$ was not calculated.}
\begin{equation}\label{mqm}
M_q=\frac{1}{\sqrt{3}}\widetilde M_q\, .
\end{equation}
From the chiral lagrangian we have
\begin{equation}
(m^2_\pi)_{ab}=2B_0\Tr\left[ M_q T_aT_b\right]\, ,
\end{equation}
that for  $m_{\pi^0}\simeq 135$ MeV and $m_{K^0}\simeq 498$ MeV gives
\begin{equation}
\label{masses}
m_u+m_d = 11.5\ {\rm MeV}\ ,\qquad
m_s  =  150\ {\rm MeV}\, .
\end{equation}
The value of the quark masses in Eq.~(\ref{masses})  are scale independent.
This is because we took  $M^2_\Phi=-3/L^2$ that corresponds, by the AdS/CFT dictionary,
 to fix the dimension of $M_q$   to be exactly one. 
In QCD however the quark masses evolve with the energy scale $\mu$.
To minimize this  discrepancy  we must  
compare   our  predictions 
with the experimental  values  of the quark masses
taken at the lowest  energy scale ($\sim 1$ GeV).
From Ref.~\cite{pdg} we have
$m_u+m_d=7-16$ MeV and $m_s=108-175$ MeV  at $\mu\sim 1$ GeV
in good agreement with  Eq.~(\ref{masses}).

\begin{figure}[t]
    \begin{minipage}[t]{0.51\linewidth}
        \begin{center}
    \psfrag{xi3}{\small{$\xi=3$}}
        \psfrag{xi4}{\small{$\xi=4$}}
        \psfrag{L5}{\small{$L_5\, [10^{-3}]$}}
        \psfrag{mS1}{\small{$M_{S_1}[{\rm GeV}]$}}
        \includegraphics[width=8.5cm]{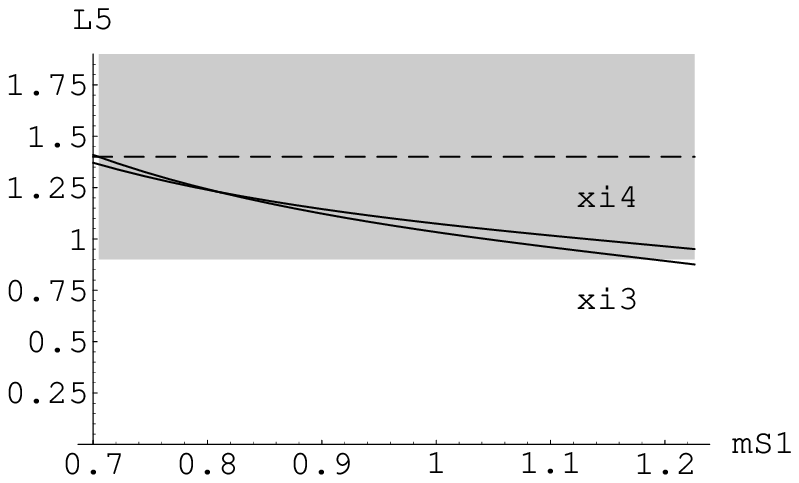}
        \end{center}
    \end{minipage}
     \begin{minipage}[t]{0.57\linewidth}
        \begin{center}
        \psfrag{xi3}{\small{$\xi=3$}}
        \psfrag{xi4}{\small{$\xi=4$}}
        \psfrag{L8}{\small{$L_8\, [10^{-3}]$}}
        \psfrag{mS1}{\small{}}
        \includegraphics[width=8.5cm]{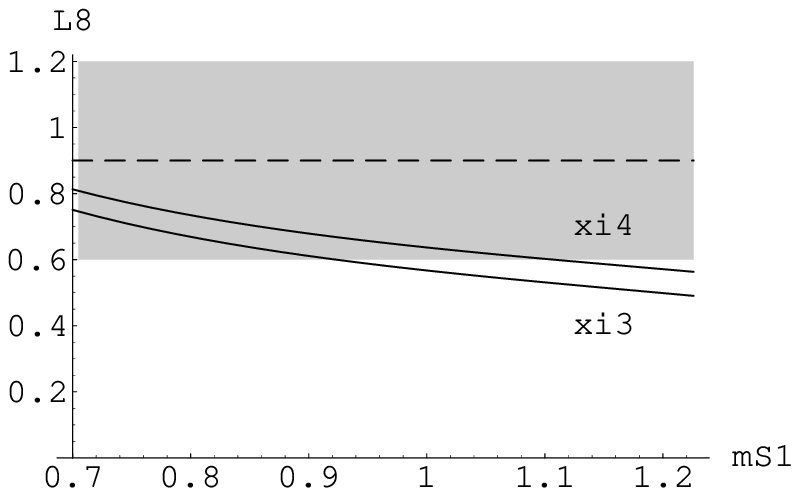}
\end{center}
    \end{minipage}
  \caption{\textit{Prediction for $L_5$ and $L_8$ as a function of $M_{S_1}$. 
  The horizontal line corresponds to the experimental value
  with the error bands \cite{Pich:1998xt}.}}
  \label{FigL8}
\end{figure}

At  ${\cal O}(p^4)$ the
chiral lagrangian  is  given by \cite{Gasser:1984gg}
\begin{eqnarray}
{\cal L}_4 &=&
L_1 \,{\rm Tr}^2\big[ D_\mu U^\dagger D^\mu U\big] +
L_2 \,\Tr\big[ D_\mu U^\dagger D_\nu U\big]
   \Tr\big[ D^\mu U^\dagger D^\nu U\big]
+ L_3 \,\Tr\big[ D_\mu U^\dagger D^\mu U D_\nu U^\dagger
D^\nu U\big]\,
\nonumber\\
&+&  L_4 \,\Tr\big[ D_\mu U^\dagger D^\mu U\big]
   \Tr\big[ U^\dagger\chi +  \chi^\dagger U \big]
+ L_5 \,\Tr\big[ D_\mu U^\dagger D^\mu U \left( U^\dagger\chi +
\chi^\dagger U \right)\big]\,
\nonumber\\
&+& L_6 \,{\rm Tr}^2\big[ U^\dagger\chi +  \chi^\dagger U \big]
+ L_7 \,{\rm Tr}^2\big[ U^\dagger\chi -  \chi^\dagger U \big]
+  L_8 \,\Tr\big[\chi^\dagger U \chi^\dagger U
+ U^\dagger\chi U^\dagger\chi\big]
\nonumber \\
&-& i L_9 \,\Tr\big[ F_R^{\mu\nu} D_\mu U D_\nu U^\dagger +
     F_L^{\mu\nu} D_\mu U^\dagger D_\nu U\big]
+  L_{10} \,\Tr\big[ U^\dagger F_R^{\mu\nu} U F_{L\mu\nu} \big]
\, .
\label{l4}
\end{eqnarray}
At tree-level, the (pseudo)scalar resonances  only contribute 
\footnote{$L_7$ will not be studied here since it arises 
from integrating out the singlet PGB that becomes massive  
when the  U(1)$_A$ anomaly  is considered.}
to $L_{1, 3,4,5,6,8}$.
The contributions to 
the coefficients $L_1$ and $L_3$ 
 coming from the octet and singlet scalar can be read from Eqs.~(\ref{L4pi8}) and (\ref{L4pi1}). 
 We obtain
 \begin{eqnarray}
L_1^{(8)}&=&-\frac{1}{3}L_3^{(8)}\,  ,\qquad
L_1^{(1)}=-L_1^{(8)} \, ,\\
L_3^{(8)}&=&
\sum_n\frac{G^2_{n\pi\pi} F^4_\pi}{8M^2_{S_n}}\, , \qquad 
L_3^{(1)}=0\, .
\end{eqnarray}
The octet and singlet contribution to the coefficient $L_1$ cancels out, as expected from 
large-$N_c$ \cite{Gasser:1984gg}, and  
  only $L_3$ gets a nonzero scalar  contribution.
For $\lambda\simeq 10^{-3}$ and $\xi=4$ ($3$)  we obtain $L^{(8)}_3\simeq 0.2\cdot 10^{-3}$ ($0.3\cdot 10^{-3}$).
Adding the vector contribution to $L_3$ calculated in Ref.~\cite{us} we get
  $L_3\simeq -2.4\cdot 10^{-3}$  ($-1.7\cdot 10^{-3}$) to be compared with  the experimental value \cite{Pich:1998xt}
   $L_3^{\rm exp}\simeq -3.5\pm 1.1$.
The scalar contribution to $L_{4}$ and $L_{5}$ can be obtained from Eq.~(\ref{sff8}):
 \begin{eqnarray}
L_4^{(8)}&=&-\frac{1}{3}L_5^{(8)}\, , \qquad L_4^{(1)}=-L_4^{(8)}\, ,\\
L_5^{(8)}&=&\frac{F^2_\pi}{8B_0}\sum_n\frac{G_{n\pi\pi}F_{S_n}}{M_{S_n}}\,  ,\qquad
L_5^{(1)}=0\, .\label{l5}
\end{eqnarray}
As expected from large-$N_c$, the total contribution to $L_4$ is zero.
The value of $L_5$ is shown in Fig.~\ref{FigL8} as a function of $M_{S_1}$ for $\xi=3,4$.
For $M_{S_1}\sim 1$ GeV we obtain $L_5\simeq 1.1\cdot 10^{-3}$ in good agreement with 
experiments.
$L_5$ can also be calculated from the axial-vector correlator \cite{us}:
\begin{equation}
\label{ l5b}
L_5=\left.\frac{1}{16B_0}\frac{d\Pi_A}{d M_q}\right|_{M_q=0}\, .
\end{equation}
For $\xi\gg 1$ with  $\lambda\xi^2$ fixed \footnote{In Ref.~\cite{us} the value of $L_5$ was given for 
$\lambda\xi^2\gg 1$ and therefore the last term of Eq.~(\ref{l5ap}) was not present.
This last term arises due to the $\xi$ dependence on $\widetilde M_q$ - see Eq.~(\ref{xi}).
Also a factor $1/2$ was missing in Eq.~(67)  of Ref.~\cite{us} and therefore the 
 prediction of $L_5$ given there was a factor $2$ larger.},
 we obtain
\begin{equation}
\label{l5ap}
L_5\simeq
\frac{\tilde N_c\pi^3}{\sqrt{3}
\Gamma(\frac{1}{3})^6}\left[1-\frac{2\tilde N_c}{3F^2_\pi L_1^2}\right]+
\frac{F^4_\pi L_1^4}{192\lambda\xi^4}
\simeq 1.2\cdot 10^{-3}\left[1-0.23\left(\frac{4}{\xi}\right)^{\frac{2}{3}} 
+0.09\left(\frac{10^{-3}}{\lambda}\right)
\left(\frac{4}{ \xi}\right)^{\frac{8}{3}} 
\right]\, .
\end{equation}

Finally,  the coefficient $L_{6,8}$ can be computed from the correlators $\Pi_{S,P}$. We have
 \begin{eqnarray}
L_6^{(8)}&=&-\frac{1}{3}L_8^{(8)}\, , \qquad L_6^{(1)}=-L_6^{(8)}\, ,\\
L_8^{(8)}&=&\left.\frac{1}{32B_0^2} \frac{d}{dp^2}\Big[p^2\big(\Pi_S(p^2)-\Pi_P(p^2)\big)\Big]\right|_{p^2=0}
\,  ,\qquad
L_8^{(1)}=0\, .
\end{eqnarray}
Then $L_6=L_6^{(8)}+L_6^{(1)}=0$, as expected from large-$N_c$. 
Using  Eqs.~(\ref{smallpS}), (\ref{smallpP}) and (\ref{b0}) in the above equation,  
we obtain 
\begin{equation}
L_8\simeq\frac{\tilde{N}_c}{32}
\left[1-\frac{6}{B_0^2L_1^2}+\frac{3\tilde{N}_c}{2\lambda \xi^2B_0^2L_1^2}\right]
  \simeq
8\cdot 10^{-4}\left[1
-0.27\left(\frac{4}{\xi}\right)^{\frac{2}{3}}
+0.11\left(\frac{10^{-3}}{\lambda}\right)
\left(\frac{4}{ \xi}\right)^{\frac{8}{3}} 
\right]\, .
\end{equation}
Notice that this expression is only valid for $\xi\gg 1$ with $\lambda\xi^2$ fixed.
In Fig.~\ref{FigL8} we show the exact value of $L_8$ as a function of $M_{S_1}$. 
For $M_{S_1}\simeq 1$ GeV and $\xi=4$ we obtain $L_8\simeq 0.6\cdot 10^{-3}$ 
again in good agreement with the  experimental value. 
From Fig.~\ref{FigL8}  one can see that  small values of  $M_{S_1}$ are preferred.  
The coefficient $L_8$ can also be written as 
\begin{equation}
\label{l8kk}
L_8
=\frac{1}{32B_0^2}\left[F_{S_1}^2+\sum_{n=1}^\infty\Big(F^2_{S_{n+1}}-F^2_{P_n}\Big)\right]
\, ,
\end{equation}
that  shows that in the limit where the  chiral symmetry is restored,   
 $\xi\rightarrow 0$ and $F_{S_{n+1}}\rightarrow F_{P_n}$,  only the first term remains.
For $\xi\simeq 4$ 
 we  find that the first term  still  dominates  (it gives $70\% $ of the total  contribution
for $\lambda\simeq 10^{-3}$)
since the other resonances, being so heavy, are not very sensitive to chiral
symmetry breaking.

\section{Conclusions}

We have analyzed  the scalar and pseudoscalar sector of 
a five-dimensional model proposed to study 
mesons in  QCD. 
  We have calculated the scalar and pseudoscalar two-point correlator
and we have obtained the  mass spectrum and interactions.
This has allowed us to
determine the (pseudo)scalar   contribution to the scalar form factor of the 
PGB as well as the contribution to the $L_i$ coefficients of the chiral lagrangian.
We have also found   two interesting sum rules  for the scalar couplings and masses of the resonances that are fulfilled 
 generically    in AdS$_5$ models.

Comparing with the experimental data, we have found a good agreement for the 
$L_i$ predictions (see Fig.~\ref{FigL8}) and the  quark masses.
For the  first massive pseudoscalar resonance we have obtained  
 a mass   around $1800$ MeV,
quite different from the mass of the lowest QCD pseudoscalar  resonance $\pi(1300)$.
This has suggested us to associate this state to  $\pi(1800)$.
We have also given  predictions for the scalar couplings and decay constants
but the absence of clean experimental data  has not allowed us to compare them with QCD.

Previous approaches to calculate the scalar and pseudoscalar spectrum and/or
determine their contribution to $L_i$  can be found in 
Refs.~\cite{Peris:1998nj}-\cite{Bramon:2003xq}.
In particular, the  analysis of Refs.~\cite{Golterman:1999au,Jamin:2001zq} 
has certain similarity with ours.   
Refs.~\cite{Golterman:1999au,Jamin:2001zq}
   work in the   large-$N_c$  limit
where   QCD is described as  a theory of infinite  hadron resonances.
These   sets of infinite  hadrons, however,   are approximated 
in Refs.~\cite{Golterman:1999au,Jamin:2001zq}
by   taking only the lowest  modes, and
their masses and couplings   are determined
by  demanding  a good high-energy   
behaviour of the  correlators and form factors.
In our approach we have shown that the correlators and  form factors  have   the correct
high-energy behaviour since
  this   is dictated by the conformal symmetry.
We have also found that, in certain cases,  it can be a good  approximation to
take only the  lowest  resonance.  
Therefore in  these cases  our approach and that of 
Refs.~\cite{Golterman:1999au,Jamin:2001zq}   give similar results.
 Nevertheless, we have showed that the single-resonance approximation 
is not always  justified  (for example in Eq.~(\ref{sumrule2}))
and  this approximation can lead to large errors  in the determination of the scalar
parameters.

The analysis carried out here can be  extended to  study three-point or 
four-point correlation functions
 or to incorporate the 
 effects of $m_s$ either in the mass spectrum 
or in the interactions.    Also the effects of  higher-dimensional operators 
or  departures from AdS$_5$ in the IR-boundary  can be studied.
 These effects are important
 to study  the power corrections in the correlators.
We leave this analysis for the future.

\section*{Acknowledgements}
\label{sec:acknowledge}

We would like to thank Rafel Escribano, Matthias Jamin, Santi Peris 
and Eduard de Rafael for very useful discussions. 
This work  was  partly supported   by the MCyT and
FEDER Research Project FPA2002-00748 and DURSI Research Project
2001-SGR-00188.
The work of LD was supported 
by the Spanish Education Office (MECD) under 
an FPU scholarship.

\end{document}